\documentclass[10pt, conference, letterpaper]{IEEEtran}

\IEEEoverridecommandlockouts
\IEEEpubid{\makebox[\columnwidth]{978-3-903176-27-0~\copyright2020 IFIP \hfill} \hspace{\columnsep}\makebox[\columnwidth]{ }}

\usepackage{adjustbox}
\usepackage{algorithmic}
\usepackage{amsfonts}
\usepackage{amsmath}
\usepackage{amssymb}
\usepackage[american]{babel}
\usepackage{booktabs}
\usepackage{chronosys}
\usepackage{epsfig}
\usepackage{graphicx}
\usepackage{hyperref}
\usepackage{cleveref}
\usepackage{listings}

\usepackage{paralist}
\usepackage{pifont}
\usepackage{siunitx}
\usepackage{subfigure}
\usepackage{tabularx}
\usepackage{textcomp}
\usepackage{tikz}
\usepackage{wasysym}
\usepackage{xcolor}
\usepackage{xspace}

\hypersetup{hidelinks=true,breaklinks=true}
\usetikzlibrary{timeline}

\definecolor{base03}{HTML}{002B36}
\definecolor{base02}{HTML}{073642}
\definecolor{base01}{HTML}{586E75}
\definecolor{base00}{HTML}{657B83}
\definecolor{base0}{HTML}{839496}
\definecolor{base1}{HTML}{93A1A1}
\definecolor{base2}{HTML}{EEE8D5}
\definecolor{base3}{HTML}{FDF6E3}
\definecolor{yellow}{HTML}{B58900}
\definecolor{orange}{HTML}{CB4B16}
\definecolor{red}{HTML}{DC322F}
\definecolor{magenta}{HTML}{D33682}
\definecolor{violet}{HTML}{6C71C4}
\definecolor{blue}{HTML}{268BD2}
\definecolor{cyan}{HTML}{2AA198}
\definecolor{green}{HTML}{859900}

\definecolor{darkgreen}{HTML}{237a06}
\definecolor{darkred}{HTML}{bc2e0b}

\lstdefinelanguage{JavaScript}{
  keywords={typeof, new, true, false, catch, function, return, null, catch, switch, var, if, in, while, do, else, case, break, for},
  keywordstyle=\color{blue}\bfseries,
  ndkeywords={class, export, boolean, throw, implements, import, this},
  ndkeywordstyle=\color{darkgray}\bfseries,
  identifierstyle=\color{black},
  sensitive=false,
  comment=[l]{//},
  morecomment=[s]{/*}{*/},
  commentstyle=\color{purple}\ttfamily,
  stringstyle=\color{red}\ttfamily,
  morestring=[b]',
  morestring=[b]"
}

\newcommand{\eg}{{e.g.,}\xspace}
\newcommand{\ie}{{\it i.e.,}\xspace} 

\newif\iftodo

\newif\ifcomment
\commenttrue

\ifcomment
        \newcounter{MVNumberOfComments}
    \stepcounter{MVNumberOfComments}
    \newcommand{\mvnote}[1]{\textcolor{blue}{\small \bf [MV\#\arabic{MVNumberOfComments}\stepcounter{MVNumberOfComments}: #1]}}
    
      \newcounter{DPNumberOfComments}
    \stepcounter{DPNumberOfComments}
    \newcommand{\dpnote}[1]{\textcolor{magenta}{\small \bf [DP\#\arabic{DPNumberOfComments}\stepcounter{DPNumberOfComments}: #1]}}
\else
    \newcommand\mvnote[1]{}
     \newcommand\dpnote[1]{}
\fi

\newcommand{\etal}[0]{\mbox{\textit{et al.}}\xspace}

\newcommand{\js}[0]{JavaScript\xspace}
\newcommand{\coinhive}[0]{Coinhive\xspace}
\newcommand{\authedmine}[0]{Authedmine\xspace}
\newcommand{\jsecoin}[0]{JSECoin\xspace}
\newcommand{\cryptonight}[0]{Cryptonight\xspace}

\newcommand{\mysubpara}[1]{\vskip 0.1cm \noindent\textit{#1}.\xspace}

\def\BibTeX{{\rm B\kern-.05em{\sc i\kern-.025em b}\kern-.08em
    T\kern-.1667em\lower.7ex\hbox{E}\kern-.125emX}}

\newcommand{\cmscanner}[0]{\textit{cm-screener}\xspace}

\newcommand{\vt}[1]{\rotatebox{90}{#1}}

\todotrue
\commenttrue

\makeatletter
\newcommand{\linebreakand}{  \end{@IEEEauthorhalign}
  \hfill\mbox{}\par
  \mbox{}\hfill\begin{@IEEEauthorhalign}
}
\makeatother

\begin{document}
\title{A Retrospective Analysis of User Exposure to (Illicit) Cryptocurrency Mining on the Web}

\author{\IEEEauthorblockN{Ralph Holz}
\IEEEauthorblockA{\textit{University of Twente \& University of Sydney}}
\and
\IEEEauthorblockN{Diego Perino}
\IEEEauthorblockA{\textit{Telefonica Research}}
\and
\IEEEauthorblockN{Matteo Varvello}
\IEEEauthorblockA{\textit{Brave Software}}
\and
\IEEEauthorblockN{Johanna Amann}
\IEEEauthorblockA{\textit{ICSI}}
\linebreakand
\IEEEauthorblockN{Andrea Continella}
\IEEEauthorblockA{\textit{University of Twente}}
\and
\IEEEauthorblockN{Nate Evans}
\IEEEauthorblockA{\textit{University of Denver}}
\and
\IEEEauthorblockN{Ilias Leontiadis}
\IEEEauthorblockA{\textit{Samsung AI}}
\and
\IEEEauthorblockN{Christopher Natoli}
\IEEEauthorblockA{\textit{University of Sydney}}
\and
\IEEEauthorblockN{Quirin Scheitle}
\IEEEauthorblockA{\textit{Technical University of Munich}}
}

\maketitle

\begin{abstract}

    In late 2017, a sudden proliferation of malicious \js was reported on the
    Web: browser-based mining exploited the CPU time of website visitors to
    mine the cryptocurrency Monero. Several studies measured the deployment of
    such code and developed defenses. However, previous work did not establish
    how many users were really \textit{exposed} to the identified mining sites
    and whether there was a real risk given common user browsing behavior. In
    this paper, we present a retroactive analysis to close this research gap.
    We pool large-scale, longitudinal data from several vantage points,
    gathered during the prime time of illicit cryptomining, to measure the
    impact on web users. We leverage data from passive traffic monitoring of
    university networks and a large European ISP, with suspected mining sites
    identified in previous active scans. We corroborate our results with data
    from a browser extension with a large user base that tracks site visits. We
    also monitor open HTTP proxies and the Tor network for malicious injection
    of code. We find that the risk for most Web users was always very low, much
    lower than what deployment scans suggested. Any exposure period was also very
    brief. However, we also identify a previously unknown and exploited attack
    vector on mobile devices.

\end{abstract}

\section{Introduction}

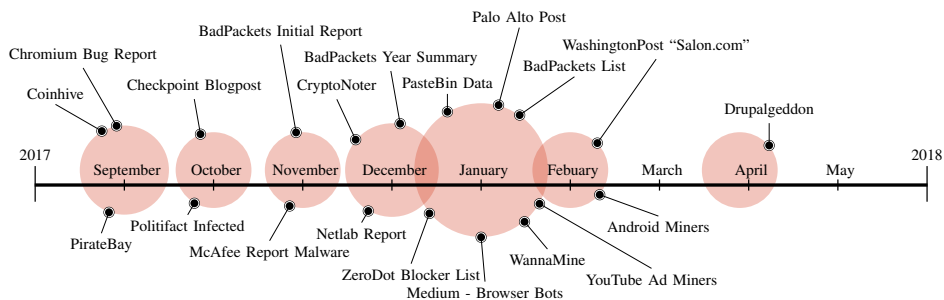
\begin{figure*}[t]
                              \begin{center}
      \resizebox{.7\textwidth}{!}{
      \begin{tikzpicture}[timespan={}]

        \timeline[custom interval=true]{September, October, November, December, January, Febuary, March, April, May}

\begin{phases}
  \phase{between week=1 and 1 in 0.1,involvement degree=2.0cm}
\phase{between week=2 and 2 in 0.2,involvement degree=1.7cm}
\phase{between week=3 and 3 in 0.3,involvement degree=1.7cm}
\phase{between week=4 and 4 in 0.4,involvement degree=2.1cm}
\phase{between week=5 and 5 in 0.5,involvement degree=3.0cm}
\phase{between week=6 and 6 in 0.6,involvement degree=1.7cm}
\phase{between week=7 and 8 in 0.9,involvement degree=1.7cm}

\end{phases}

\addmilestone{at=phase-1.120,direction=150:1.2cm,text={\coinhive},text options={above}}
\addmilestone{at=phase-1.250,direction=250:0.5cm,text={PirateBay},text options={below}}
\addmilestone{at=phase-1.100,direction=120:1.5cm,text={Chromium Bug Report},text options={above}}
\addmilestone{at=phase-2.240,direction=240:0.3cm,text={Politifact Infected},text options={below}}
\addmilestone{at=phase-2.110,direction=100:0.8cm,text={Checkpoint Blogpost},text options={above}}
\addmilestone{at=phase-3.100,direction=100:2.0cm,text={BadPackets Initial Report},text options={above}}
\addmilestone{at=phase-3.250,direction=240:0.9cm,text={McAfee Report Malware},text options={below}}
\addmilestone{at=phase-4.140,direction=115:1.0cm,text={CryptoNoter},text options={above}}
\addmilestone{at=phase-4.240,direction=240:0.3cm,text={Netlab Report},text options={below}}
\addmilestone{at=phase-4.80,direction=100:1.2cm,text={BadPackets Year Summary},text options={above}}
\addmilestone{at=phase-5.220,direction=250:1.2cm,text={ZeroDot Blocker List},text options={below}}
\addmilestone{at=phase-5.270,direction=275:1.0cm,text={Medium - Browser Bots},text options={below}}
\addmilestone{at=phase-5.120,direction=90:0.4cm,text={PasteBin Data},text options={above}}
\addmilestone{at=phase-5.75,direction=75:1.8cm,text={Palo Alto Post},text options={above}}
\addmilestone{at=phase-5.55,direction=35:1.5cm,text={BadPackets List},text options={above}}
\addmilestone{at=phase-5.310,direction=310:0.8cm,text={WannaMine},text options={below}}
\addmilestone{at=phase-5.330,direction=330:2.9cm,text={YouTube Ad Miners},text options={below}}
\addmilestone{at=phase-6.320,direction=340:1.4cm,text={Android Miners},text options={below}}
\addmilestone{at=phase-6.45,direction=55:2.3cm,text={WashingtonPost ``Salon.com''},text options={above}}
\addmilestone{at=phase-7.40,direction=90:0.5cm,text={Drupalgeddon}, text options={above}}

\node (2017) at (0,1) {2017};
\path[-, thick] (2017) edge (0, 0.2);
\node (2018) at (20,1) {2018};
\path[-, thick] (2018) edge (20, 0.2);

      \end{tikzpicture}
    }
   \end{center}
   \vskip -.5 cm
   \caption{Timeline of key events during the rise of cryptojacking. Circle size is a rough measure of the number of reports in the respective period.\label{fig:timeline}}
 \end{figure*}

In September 2017, the filesharing site Piratebay was reported to engage in
browser-based mining of the Monero cryptocurrency by including \js and Web
Assembly (WASM) code from the \coinhive mining pool. Due to Monero's
cryptographic design, such mining was viewed as a possible revenue model for
site operators. Dozens of pool operators and mining variants emerged soon (see
\Cref{fig:timeline}). The term cryptojacking was soon coined for sites that
exploited their visitors' CPU without informing them; attackers even
compromised websites ~\cite{washigtonpost-politifact} and exploited known
weaknesses of the Drupal web framework~\cite{Drupalgeddon2018,
Drupalgeddon2018a} to plant mining code.  Starting in late 2018, a number of
researchers investigated the deployment of cryptomining code on websites and
the operator ecosystem~\cite{Rueth2018,Konoth2018,Hong2018,Kharraz2019}. While
their results are not quite consistent, they still demonstrated that deployment
was occurring on thousands of websites or more.

These previous lines of research left some important questions unanswered,
however, as the methods used were almost exclusively short-term, active scans
of parts of the web. These are well-suited to gauging deployment, but they
cannot establish if users were actually \textit{exposed} to mining as they do
not measure \textit{how many} and \textit{how often} users encounter mining
sites. This requires \emph{passive} observation of user behavior to understand
the actual risks. The above methods also miss out on attack vectors that
are known to be of practical relevance, such as the injection of mining code by
HTTP proxies~\cite{proxytorrent}.

In this paper, we pool existing datasets from several research groups and
vantage points that were created during 2018-2019. We present a
\emph{retrospective}, longitudinal view of user exposure to browser-based
cryptocurrency mining and investigate various forms of exposure due to the
injection of mining code by intermediary nodes. Our main contributions are as
follows:

\mysubpara{1) User exposure} We measure user exposure by leveraging
passive traffic monitoring, using monitoring stations in several North American
research and education networks, a large European mobile ISP and an existing
Chrome extension with a large user base. Our measurements are longitudinal. We
show that the problem was by far not as often experienced by users as previous
deployment measurement suggested: users were rarely, if at all, ever exposed.

\mysubpara{2) Alternative attack vectors} We check the ecosystem of open
proxies for malicious injection of cryptomining code, testing up to 250k
proxies per day, and monitor exit nodes on the Tor network for injections. We
show that the attack vector was real and exploited, but fortunately most users
were not at risk.

\mysubpara{3) Unknown attack vector} We identify an unknown attack
vector in our passive data and trace it to manufacturers and/or suppliers of
cheap mobile phones who upload apps without user consent. The apps cause
cryptocurrency to be mined on the phones.

The remainder of this paper is organized as follows. We give an overview of
related work and position our contribution in \Cref{sec:relwork}.
\Cref{sec:methodology} presents our methodology and discusses ethical
considerations. We present our results in \Cref{sec:results} and discuss them
in \Cref{sec:discussion}.

\section{Related work}
\label{sec:relwork}

Previous work on in-browser mining focused nearly exclusively on deployment,
campaigns, defenses, and estimates of revenue. To the best of our knowledge,
passively obtained data is only used in \cite{Bijmans2019} in a 30,000-ft
estimate to show that some users were affected; however, this is not put into
context of the overall user base. Injection of mining code by proxies, Tor
nodes or mobile phone suppliers has not received any attention. Hence, our work
is complementary to previous work.  Table~\ref{tab:relwork} visualizes this and
summarizes previous contributions. 

\begin{table*}
    \centering
    \caption{Comparison with related work. Half circles refer to methodological limitations documented in each research paper.}
    \label{tab:relwork}
    \scalebox{0.8}{
    \begin{tabular}{p{2.45cm}llllll|llp{1.6cm}p{2.4cm}ll|lll|lllll|c}
        \toprule
        ~   & \multicolumn{6}{c}{\textbf{Contributions}} & \multicolumn{6}{c}{\textbf{Methodology}}    & \multicolumn{8}{c}{\textbf{Classifiers in scans}} & \textbf{Incidence \%} \\
        ~   & \multicolumn{6}{c}{~}                      & \multicolumn{6}{c}{~}                                & \multicolumn{3}{c}{\textbf{static}} & \multicolumn{5}{c}{\textbf{dynamic}} & ~ \\
        \midrule

        ~ & \vt{User exposure} & \vt{Conn. injection}  & \vt{Deployment} & \vt{Campaigns} & \vt{Revenue} & \vt{Mitigation}
        & \vt{Monitoring} & \vt{Active scans} & \vt{Scan targets}  & \vt{Period} & \vt{Links followed}  & \vt{Longitudinal} 
        & \vt{Patterns (samples)}   & \vt{Patterns (blocklists)}  & \vt{Signatures}   
        & \vt{CPU usage} &   \vt{CPU cache} & \vt{Code monitoring}  & \vt{Network stack} & \vt{Library detection} 
        & ~ \\

        \midrule

        Bijmans \etal \cite{Bijmans2019} & \LEFTcircle & \Circle & \CIRCLE & \CIRCLE & \CIRCLE & \Circle
        & \LEFTcircle & \CIRCLE & Top1m; 49m & 2018-12--2019-01 & \Circle & \Circle
        & \Circle & \CIRCLE & \Circle
        & \LEFTcircle & \Circle & \CIRCLE & \CIRCLE & \Circle & 0.01-0.07\% \\
        \addlinespace[.1cm]

        Hong \etal \cite{Hong2018}    & \Circle  & \Circle & \CIRCLE & \CIRCLE & \CIRCLE & \Circle 
        & \Circle & \CIRCLE & Top100k & 2018-04  & \CIRCLE & \Circle 
        & \Circle & \Circle & \Circle  
        & \Circle  & \Circle  & \CIRCLE  & \Circle & \Circle & 0.50-0.87\% \\
        \addlinespace[.1cm]

        Kharraz \etal \cite{Kharraz2019} & \Circle & \Circle & \CIRCLE & \CIRCLE & \Circle & \Circle 
        & \Circle & \CIRCLE & Top1m; 600k & 2018-02--2018-10 & \CIRCLE & \CIRCLE 
        & \Circle & \Circle & \Circle 
        & \Circle & \Circle & \CIRCLE & \CIRCLE & \Circle & 0.59\% \\
        \addlinespace[.1cm]

        Konoth \etal \cite{Konoth2018} & \Circle & \Circle & \CIRCLE & \CIRCLE & \CIRCLE & \CIRCLE 
        & \Circle & \CIRCLE & Top1m & 2018-02 & \CIRCLE & \Circle 
        & \Circle & \CIRCLE & \Circle 
        & \LEFTcircle & \CIRCLE & \CIRCLE & \CIRCLE & \Circle & 0.07-0.17\% \\
        \addlinespace[.1cm]

        R\"{u}th \etal \cite{Rueth2018}  & \Circle & \Circle & \CIRCLE & \LEFTcircle & \CIRCLE & \Circle 
        & \Circle & \CIRCLE & Top1m; 12m & 2018-01--2018-05 & \Circle & \CIRCLE 
        & \Circle & \CIRCLE & \CIRCLE 
        & \Circle & \Circle & \Circle & \CIRCLE & \Circle & 0.08\% \\
        
        \midrule

        \textbf{This work} & \CIRCLE & \CIRCLE & \LEFTcircle & \Circle & \Circle & \Circle 
        & \CIRCLE & \CIRCLE & Top1m; 265m & 2018-01--2018-11 & \Circle & \CIRCLE 
        & \CIRCLE & \CIRCLE & \CIRCLE 
        & \Circle & \Circle & \Circle & \Circle & \CIRCLE & 0.02-0.12\% \\
        \bottomrule
    \end{tabular}
}
\end{table*}
\normalsize

Previous work used various active detection methods and counted incidents quite
differently. This makes it very difficult to compare between the authors. Some
authors investigated only the landing pages of websites~\cite{Rueth2018};
others followed internal links \cite{Konoth2018,Bijmans2019}; others even
counted links to external sites as a mining occurrence \cite{Hong2018}. Some
authors used static detection and pattern matching in their scans
\cite{Rueth2018, Bijmans2019}, others used dynamic code analysis and/or
hardware monitoring \cite{Konoth2018,Kharraz2019}.  Many investigated only
domains from Alexa Top1m lists on single or very few occasions, although the
high variability of this data source (up to 50\% churn per day) is well
established~\cite{toplistsimc18}. Others again sampled a large part of domains
from public DNS zonefiles~\cite{Rueth2018,Bijmans2019}. It is usually
impossible to give a reliable estimate how many users visited the respectively
identified mining sites.

The authors of \cite{Hong2018,Kharraz2019,Konoth2018,Rueth2018} carried out
active scans of Alexa lists. Detection is mostly based on dynamic analysis:
monitoring the creation of Websockets, use of the Stratum protocol, and
profiling the call stack and \js worker threads. These methods have high
accuracy. However, they are not applicable in a passive measurement because the
affected machine can only be monitored in terms of network traffic. The
publications most closely related to ours are probably the ones by Konoth \etal
\cite{Konoth2018} and Bijmans \etal \cite{Bijmans2019}, which use the same
toolset (using dynamic detection of the use of Websockets and the protocol to
communicate with a mining pool). Bijmans~\etal also analyze the nature and
purpose of mining sites; this was out of scope for our study. Konoth \etal find
that static string matching is already sufficient to detect 93\% of mining
sites. Dynamic detection is useful to eliminate false positives, however. They
also propose a defense based on measuring the CPU cache and identifying
cryptographic primitives in the call stack. They determine an incidence rate of
0.07\% for landing pages and 0.17\% when including internal links. Bijmans
\etal report 0.01\% in a smaller scan but 0.07\% for domains on the Alexa list,
concluding that Alexa scans overestimate the incidence rate. Hong \etal and
Kharraz \etal's tools produce deviating results (0.50\%-0.87\% incidence), but
they count links to external sites as an incidence relating to the investigated
site. Rueth \etal's study~\cite{Rueth2018} is quite different: it uses pattern
matching for detection and dissects the Coinhive link-forwarding service.

\section{Methodology}
\label{sec:methodology}

We use passive measurements to obtain an accurate picture of user exposure.
Some of our passive measurements rely on input from active scans and
classifications from blocklists to identify mining sites.  We support open
science and make code and non-privacy sensitive data in this paper available:
\begin{center}
\url{https://github.com/retrocryptomining}
\end{center}

\begin{table}
    \centering
        \caption{Our samples by family and origin. Helper code counted across families.}
        \label{tab:samples}
        \scalebox{0.8}{
        \begin{tabular}{lrp{4.1cm}}
            \toprule
            \textbf{Family/type}    & \textbf{Samples (obfuscated)}  & \textbf{Sources}     \\
            \midrule
            \multicolumn{3}{l}{Cryptonight family\ldots} \\
            ~~Authedmine              & 6                 (6)                                              & \textit{authedmine.com}; affected sites \\
            ~~Coinhive                & 9                 (0)                                              & \textit{coinhive.com}, \textit{npm}; affected sites \\
            ~~Cryptoloot              & 10                (8)                                              & \textit{Cryptoloot} (Github); affected sites \\
            ~~Other & 37                (27)                                             &  respective pools; affected sites \\
            \midrule
            JSECoin                 & 2                 (0)                                              & \textit{JSECoin}; affected site \\
            \midrule
            Helper code             & 14                (1)                                              & respective pools; affected sites \\
        \bottomrule
        \end{tabular}
    }
\end{table}

\subsection{Deriving Classifiers}
\label{sec:sub:samples}

\mysubpara{Samples} Mining code consists of two parts: a part responsible for the
mining (WASM or, rarely, custom encodings) and helper code for configuration.
We collected both helper and mining code, including for those mining pools that
require manual user opt-in. We collected 78 samples between 2017-10 and
2018-07, from affected sites and by creating accounts with mining pools. This
was a manual, iterative process following reports on the Web
(especially~\cite{badpacketstwitter}). We summarize our collection in Table
\ref{tab:samples}.  In line with previous work, we find two algorithms
dominate: \cryptonight and JSECoin. Mining pools use different \cryptonight
implementations; however a significant number of samples are just variants of
the implementations by \coinhive, \authedmine, and CryptoLoot.  Some
implementations use a custom byte encoding, not WASM. During our collection
period, we noted a shift to heavy obfuscation, especially in the more recent
\cryptonight implementations by CoinNebula and DeepMiner.
We apply deobfuscation to verify a sample is mining code. For \authedmine, we
develop a custom reverser.  DeepMiner encrypts its code with AES; the key is
located in the \js itself. For other implementations, we apply partial
evaluation using the open-source tool JStillery. Owing to our long collection
period and the fact that most implementations are closely related variants of
\cryptonight, we believe our samples provide excellent coverage of most
cryptocurrency mining in use in 2018-2019.

\mysubpara{Blocklists} We use classifications from five important blocklists as
further input. All are available on Github or via the respective browser
extension: NoCoin, coinhive-block, coinhive-blocker, crypto-miner-blocker, and
MinerBlock.  We consolidate these from 2017-11 to 2018-07. We eliminate
overambitious regular expressions (\eg single-letter filenames). The regular
expressions we deem suitable for detection cover URLs of mining domains, URLs
of Websockets, \js filenames, and a small number of highly specific keywords.

\mysubpara{Further input} The authors of~\cite{Konoth2018} provide us with a list
of mining sites they identified in 2018-02.

\label{sec:sub:classifiers}

From our samples and regular expressions, we derive a number of classifiers.
Two forms of classifiers are for static analysis in active and passive
measurements. One classifier can be used to inspect \js code at runtime.

\mysubpara{C1: pattern matching} Classifiers in this category are string
patterns for both obfuscated and non-obfuscated mining code, including various
alternative encodings of WASM. In \mbox{2018-01}, there are 45 classifiers in this
category (one was later found to be redundant). Our set grows to 58 in 2018-03 and to
364 in 2018-07. This was due to fast growth of the blocklists, which included
ever longer lists of proxy domains for Websockets, used by mining scripts to
avoid detection.

\mysubpara{C2: signatures} We implement MD5, fuzzy
hashing\footnote{github.com/DinoTools/python-ssdeep}, and Yara
signatures\footnote{github.com/VirusTotal/yara} to identify and differentiate
between variants of mining scripts.

\mysubpara{C3: namespaces} We find that global variables are commonly used in
mining code to simulate namespaces, with variable names often unique to a
family of related implementations. We use this for classification in our
browser extension.

An obvious limitation of our approach is that we can only detect known samples
and closely related derivates that preserve the key characteristics we
extracted. We argue that this limitation should be of little relevance for this
study due to our broad coverage.

\subsection{Active Scanning of Domains}
\label{sec:sub:methods-active-scans}

We use scans of landing pages to validate the coverage and accuracy of our
classifiers and create input data for our passive measurements.  Our scanner
(\cmscanner) is custom-built and written in Go.
We pick scan targets from two sources. We scan domains on the Alexa Top 1m,
with a fresh list downloaded before each scan. We also reuse the methodology
from~\cite{Scheitle2018}. We collect domains from ICANN's CZDS
(\textit{czds.icann.org}), Certificate Transparency logs, and the
\textit{.com}, \textit{.net}, \textit{.org} zones. We add domains from the
Umbrella and Majestic top lists. We resolve all domains with \textit{massdns}
and scan IP addresses with \textit{zmap} to eliminate unreachable domains.

Initally, we use \cmscanner to test landing pages for the presence of links to
\js mining code, in particular links to the respective mining pools. This was a
very common deployment pattern in the early phase of in-browser mining. Such a
scan can screen nearly 100m domains in under 24h. With deployment diversifying
and attempting to evade detection, we extend the scanner to parse landing pages
and retrieve all \js (inline and linked) and match the code against our
classifiers. The parsing is considerably slower; we limit ourselves to one full
scan of the Alexa list and just under 20m random domains from our larger input
list. We evaluate our detection rate by statistical sampling and determine the
sample proportion of true and false positives (we count commented mining code
as a false positive). We show in ~\Cref{sec:sub:results-active-scans} that this
identifies mining sites with the accuracy required for our passive
measurements.

\subsection{Passive Monitoring} \label{sec:passivecampus}

Our passive monitoring is carried out on traffic data from research and
education networks and a mobile ISP.

\mysubpara{Research and education networks} We have access to data from a long-running
project~\cite{Amann2012} monitoring TLS network connections from several North
American research and education networks with more than 100k users. This
includes campus networks and student accomodation networks. We use the Server
Name Indication (SNI) of TLS to determine the destination domain of a
connection. This allows us to identify HTTPS connections to suspected mining
domains and Websocket proxies.

Mining sites cause the browser to open Websocket connections to mining pools
and proxies. We check for these using the Websocket URLs from category C1. The
Websocket protocol is an in-band upgrade to HTTPS, \ie they are not directly
inferable. However, mining pools typically host their Websocket endpoints on
subdomains beginning with \textit{ws}, \eg \textit{ws1.coinhive.com}, and most
code samples we collect follow this pattern and connect to the proxies provided
by mining pools. Bijmans \etal independently confirm this in
\cite{Bijmans2019}; the use of other proxies is rare. Consequently, we count
only connections to domains of this format and to the pools known to us as a
mining connection.

We inspect data from our passive monitoring going back to 2015 and ending in
2018-11, 280 billion connections in total.
We pick up the first mining-related connections (to Coinhive) in 2017-08; this
was a month before the first reports.

\mysubpara{Mobile ISP} We inspect our samples and find that very few deactivate
mining on mobile browsers. Concluding that users of handheld devices should
also be affected by in-browser mining, we leverage our access to data from a
large European mobile ISP with tens of millions of subscribers in one country
to identify mining activity.  We collect passive traces and summary statistics
from 2018-01 to 2018-08. 
Specifically, we leverage two separate data sources. The first is a
transparent middlebox used by the ISP to optimize mobile traffic. The second is
a device database linking a device ID (IMEI) to a specific device model, OS,
and manufacturer. We refer to entries in our first source as `transactions'.  
Note that our methodology does \textit{not} allow us to identify tethering.

We provide the ISP with a list of regular expressions, built from our
classifiers in C1 of 2018-01. The expressions cover \coinhive, \jsecoin, and
several providers using \cryptonight. They also identify Websocket endpoints
and complete URLs of \js mining code. The ISP builds traffic capturing rules
based on these and applies them on our behalf. For HTTP, we match full URLs.
For HTTPS, we match the domain name in the SNI. We collect several TB of
transactions per day, with the following information for each matching
transaction: timestamp, anonymized subscriber ID, URL, device model, and
transaction size. Once we have obtained our statistical information, the
original data is deleted by the ISP. One limitation to note is that we cannot
update our list of regular expressions. However, all important mining providers
(e.g., \coinhive, \authedmine, and \jsecoin) are covered thanks to our
collection of samples.

\subsection{Alternative Attack Vectors: Tampering with Connections}

We explore two alternative attack vectors that have not been investigated before.

\mysubpara{Open proxies} Open web proxies relay and pseudonymize HTTP connections
free of charge. Typical use cases include hiding the geographic origin and
accessing geo-blocked content. The authors of \cite{proxyndss} and
\cite{proxytorrent} showed that such proxies may tamper with connections
and modify content to inject advertisements and malware. We test whether open
proxies inject mining code, following the methodology published in
\cite{proxytorrent}. We use an array of scanners and aggregator lists to
identify reported open proxies. Between 2017-10 and 2018-08, we access 250k
reported open proxies \textit{per day}. Of these, we classify only \num{25826}
as truly operational, \ie actually performing any proxying. We set a timeout of
60s to eliminate slow proxies; this results in \num{15892} proxies we test for
injections.  We check for content manipulation by requesting a crafted bait
webpage and comparing the received page with the original. We use classifiers
from C1 and C2 to identify the kind of injection. We \textit{manually} inspect
and verify \textit{all} results.

\mysubpara{Injection in Tor} Tor hides client IP addresses by forwarding data
through multiple routers in the network. The authors of~\cite{Winter2014}
showed that some last hops (the \emph{exit nodes}) may also tamper with
traffic. We monitor Tor exit nodes daily from 2017-12 to 2018-08. Each day, we
retrieve the list of active exit nodes from Tor directory servers. Testing each
exit node, we download static, small bait pages pages via HTTP and compare them
to versions served without Tor.
Due to the early start, we use only 11 strings as classifiers.  However, we
choose them to ascertain breadth, including keywords for \coinhive and JSECoin.
We manually verify all incidents of non-matching sites.

\subsection{Understanding User Visits to Mining Sites}

We leverage a Chrome extension deployed to investigate the ecosystem of open
proxies~\cite{proxytorrent}.  Our extension tracks connections made via
\emph{benign} open proxies. It offers users to choose a proxy from a list of
operational proxies that is continuously updated.  Proxies that modify content
are \emph{never} added to this list.  There are currently \num{7940}
installations of this extension, with about \num{1400} active users per week.
About 200GB traffic per month are inspected. To analyze whether proxy users are
affected by cryptocurrency mining activity, we instrument the extension to use
classifiers from category C1. We update the used classifiers as our list in
category C1 grows. As the extension's purpose is to prevent users from routing
their traffic via tampering proxies, any injection that is still found must
come \textit{from the visited site}. This gives us useful information about
users accessing such sites. 
In \mbox{2018-05}, we add the detection of mining libraries (C3) as the browser
extension gives us the necessary access for this form of dynamic analysis. We
analyze our dataset until 2018-08, when we deactivate the analysis
functionality.

\subsection{Ethical Considerations}

All our measurement methods have been cleared by the responsible entities and
IRBs at each participating institution. This includes all institutions
contributing data to our passive measurement of research and education
networks. The latter data collection also excludes or anonymizes sensitive
information such as client IP addresses before we see it. Data collection and
short term retention at network middle boxes of the mobile ISP is in accordance
with the terms and conditions of the ISP and the local regulations. These terms
include data processing for research and publications as allowed usage of
collected data.  We only extract aggregated information and have no access to any
personally identifiable identifiers. The Chrome extension to analyze proxy user
traffic has a clear privacy policy enabling data collection and analysis for
research and publication purposes. Analysis happens on the fly; no payloads are
collected. Only anonymous data is collected; no personally identifiable
information is included.  Our active scans follow best practices such as rate
limiting, blacklisting, and maintaining informative rDNS pointers.

\section{Results}
\label{sec:results}

We first present our results from the screening scans, which support our
passive monitoring and allow us to have confidence in our classifiers. We then
present results supporting our conclusion that very few users were exposed to
cryptocurrency mining on the web, and only for very short periods of time. We
discuss this in depth in Section~\ref{sec:discussion}.

\begin{table*}
    \footnotesize
    \begin{center}
        \caption{Screening scans. $\hat{p}$ is the sample proportion of \textit{true positives} we determined. We give a 95\% confidence interval.}
    \label{tab:scanning-datasets}
    \vspace{-1em}
    \scalebox{0.9}{
    \begin{tabular}{lllllrrrr}
        Method                     & Date                       & C1                & Input             & Linked JS     & Suspected mining sites & $\hat{p}$ (TP) & 95\%-CI $\hat{p}$ (TP)\\
        \toprule 
        \cmscanner v1                 & 31 Jan 2018              & 45                &   Top1m          & -             &   \num{1239}  & 0.96            & (0.94, 0.98) \\
        \cmscanner v1                & 12 Mar 2018               & 58                &   Top1m          & -             &   \num{994}   & 0.95            & (0.93, 0.98) \\ 
        \cmscanner v2                & 29 Jul 2018               & 364               &   Top1m          & 4.4m          &   \num{742}   & 0.78           & (0.73, 0.84) \\ 
        \midrule
        \cmscanner v1                & 14--17 Mar 2018           & 58                &   265.3m          & -             &   \num{41359} & 0.94           & (0.91, 0.97) \\
        \cmscanner v2                & 28 Jul--7 Aug 2018        & 364               &   19.7m           & 30.0m         &   \num{4398}  & 0.82            & (0.77, 0.87) \\
        \bottomrule
    \end{tabular}
}
        \vskip .1cm
    \vspace{-2em}
    \end{center}
\end{table*}

\subsection{Providing input from active scans}
\label{sec:sub:results-active-scans}

We list our screening scans in Table~\ref{tab:scanning-datasets}. For every
suspected mining domain, \cmscanner stores page dumps.\footnote{Due to a
technical issue, in a small number of cases only the mining code was stored to
disk. We used the Wayback machine to confirm it had been included in the
HTML.} We take a simple random sample of size 250 (which meets the
success/failure condition for binomial experiments) from each scan and give the
sample proportion of true positives ($\hat{p}$) together with 95\% confidence
intervals.

In 2018-01, we identify 240 of the 250 samples as true positives of sites
loading mining code that would execute in the browser. In 5 further cases, the
mining code had been commented; in five more cases our hits are false
positives. This corresponds to a true positive rate of 0.95, with a 95\%
confidence interval of (0.94, 0.98). The vast majority of the 240 cases are
inclusions of Coinhive. For the scans in \mbox{2018-03} (Alexa and large-scale
scan of 265m domains), we find very similar values. The scans in 2018-07,
however, produce a lower true positive rate.
Inspecting our results, we note an overreach in blocklist entries.  Removing
just three entries reduces the false positives by nearly half.  On the whole,
however, our incidence rates are remarkably consistent with related work. With
$\hat{p}=0.96$, the identified mining sites from the Alexa scan in 2018-01
correspond to an estimated incidence rate of 0.12. For 2018-03, we obtain 0.09.
We hence conclude that our classifiers produce results that are consistent with
other authors. We do not eliminate the false positives of 2018-07 as they are
not damaging to our passive measurements.  As we show in
\Cref{sec:sub:results-campus-networks}, user exposure is so low that checking
for too many suspected mining sites in our passive monitoring has no further
impact.

\subsection{Passive Monitoring}
\label{sec:sub:results-campus-networks}
\label{sec:sub:results-passive-monitoring}

We analyze our data from passive observations of network traffic and contrast
the results from research and education networks with those from the mobile
ISP.

\subsubsection{Research and education networks} 

We first inspect all connections in our entire observation period for
mining-related domains to which Websocket connections are made. We use the
Websocket domains/URLs from our samples and blocklists as input (C1). 
Over the entire period, we identify only 1.4m connections  (out of billions) to
a total of 398 domains. The domain names
differ only in the digits following the \textit{ws} prefix. Going through these
manually and grouping them together,
we arrive at only 15 distinct domains to which 
995k connections are made. Recall that category C1 is based on real, verified
samples and entries from blocklists, with more than 200 string patterns that
identify Websocket endpoints. Most of these receive no traffic at all.  The
amount of network traffic going to known Websocket endpoints must be described
as minuscule. The finding implies that none of these sites belong to the
popular domains on the Web that attract many users.

Grouping endpoints for mining pools and proxies together under the term
\emph{dropboxes}, we list the top domains by number of connections in
\Cref{tab:campusdb}. The \coinhive mining pools, \ie including \authedmine,
dominate by number of connections. More than 85\% of our dropbox connections
terminate there.

\begin{figure}[tb]
\centering
\includegraphics[width=0.40\textwidth]{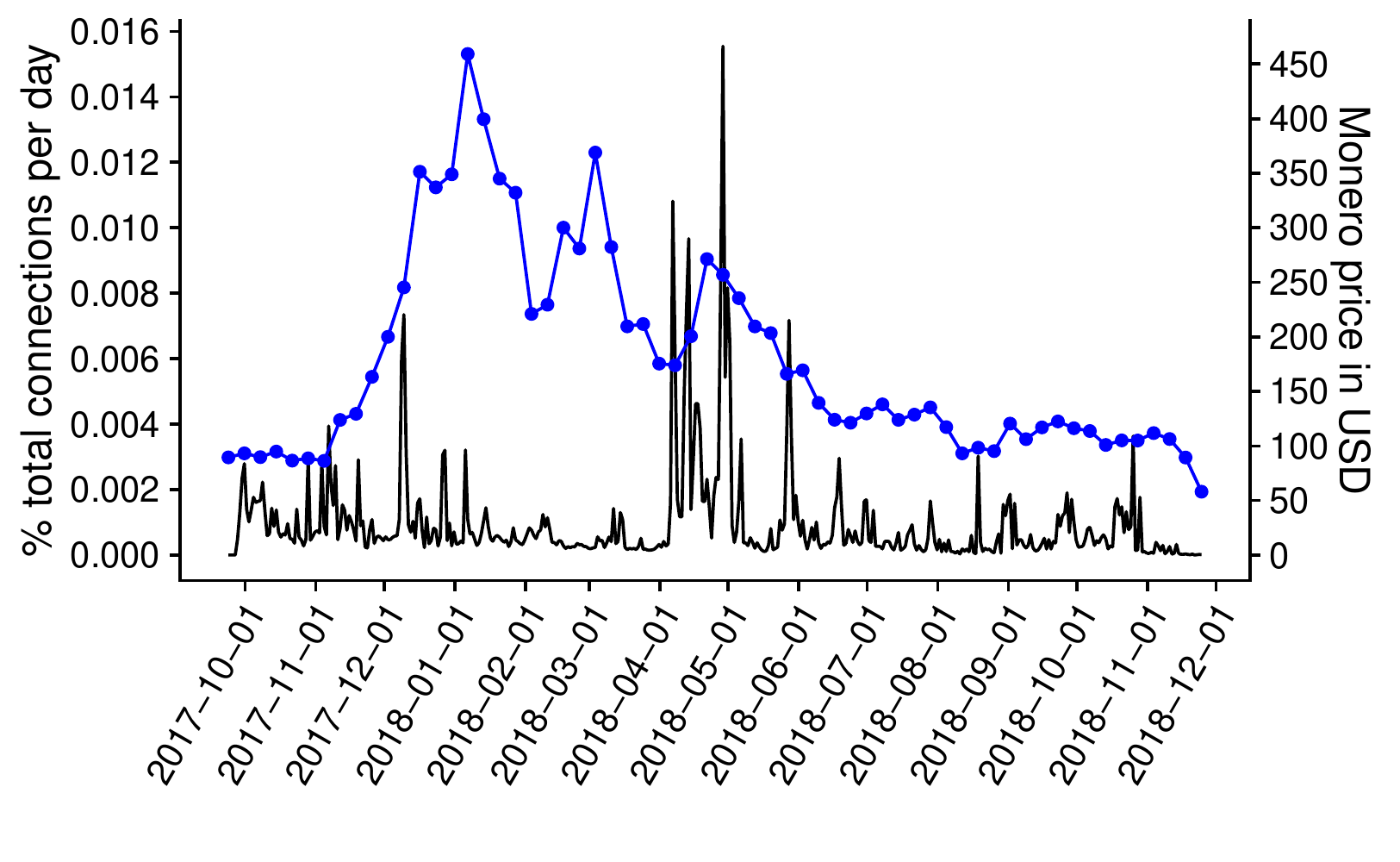}
    \caption{Black: \coinhive connections, passive measurement (\%). Blue (dotted): Monero price.}
\label{fig:coinhive-relative}
\end{figure}

\mysubpara{Coinhive} Focusing on \coinhive connections only,
\Cref{fig:coinhive-relative} shows the longitudinal development over time. The
percentage of connections per day is very low. In absolute numbers, the peaks
are at 
22k connections per day; typically, this number is much lower at several
hundred.  We observe most connections after 2017-09, when activity picks up
rapidly and peaks a first time in 2017-12, when in-browser mining was in the
news. There are two more peaks in 2018-04 and 2018-05. They coincide with the
widely reported vulnerability in the Drupal CMS (`Drupalgeddon') known to be
exploited at that time for
cryptomining~\cite{Drupalgeddon2018,Drupalgeddon2018a}. According to
\textit{coinmarketcap.com}, there was also a short price surge around this
time. We see an ever lower level of connections in the second half of the
year---at most a few thousand connections per day. We believe the reasons to
include Drupal updates, more comprehensive blocklists, and a general decline of
the cryptojacking practice due to lower prices. Note that our sample collection
is ever-growing. In 2018-07, it already contains many of the newer, alternative
mining pools.

\mysubpara{Casting a wider net} We check how many identified mining domains
appear in our scans and in our passive observations.  In the large scan of
2018-03, which covered 265m domains, we identify \num{41359} mining domains.
We search for these in our passive data, with a window reaching back 30 days.
In the 
8 billion connections of this 30-day window, we find only \num{4202}
connections to just \num{441} of the \num{41359} domains. There is a long-tail
distribution: ignoring \coinhive and \authedmine, which host the mining code
that is often linked to, we find mostly connections to movie streaming sites
(roughly 30\%). These are not the known, clearly legitimate providers but
rather sites that seem to target the `grey area' of streaming where the
illegality of consumption of such streaming is unclear, disputed, or at least
not acted upon. Just nine domains account for more than half the connections.
Almost a third of domains receive just one connection.  We repeat the analysis
for the results of the smaller scans in \mbox{2018-01} and 2018-07.  We find
\num{156} domains in \num{49742} connections (out of 
5.4 billion in the preceeding 30 days) for 2018-01. For \mbox{2018-07}, we find
\num{44} domains in 1832 connections out of a total of
2.5 billion connections in the 30-day window.

The clear picture emerging from our observations is that of a striking
discrepancy between deployment and actual user exposure. The long-tail
distribution offers evidence that blocklists in browsers and network gateways,
while definitely incomplete, can provide a much better protection than related
work assumed. We return to this in \Cref{sec:discussion}.

\begin{table}[tb]
	      \footnotesize
        \centering
        \caption{Top 10 dropboxes by connections.}
        \vspace{-0.3cm}
    \scalebox{0.8}{
        \begin{tabularx}{0.8\columnwidth}{Xr}
                \toprule
                Domain & Conns \\
                \midrule
                ws.coinhive.com & \num{760164}\\
                ws.rocks.io     & \num{83793}\\
                ws.authedmine.com       & \num{56956}\\
                ws.coin-hive.com &       \num{35068}\\
                ws.pzoifaum.info & \num{19135}\\
                ws.crypto-loot.com & \num{17683}\\
                ws.coin-have.com & \num{7548}\\
                ws.staticsfs.host & \num{3553} \\
                ws.aalbbh84.info & \num{3245}\\
                ws.hemnes.win	& \num{3060}\\
                \bottomrule
        \end{tabularx}
    }
        \label{tab:campusdb} \end{table}

\subsubsection{Mobile ISP} 
\begin{figure*}[tb]
    \begin{center}
    \subfigure[Number of unique users per day who accessed a mining domain over a period of 6 months. Cyan indicates periods where no collection was performed due to operational issues.]{
        \includegraphics[width=0.31\textwidth]{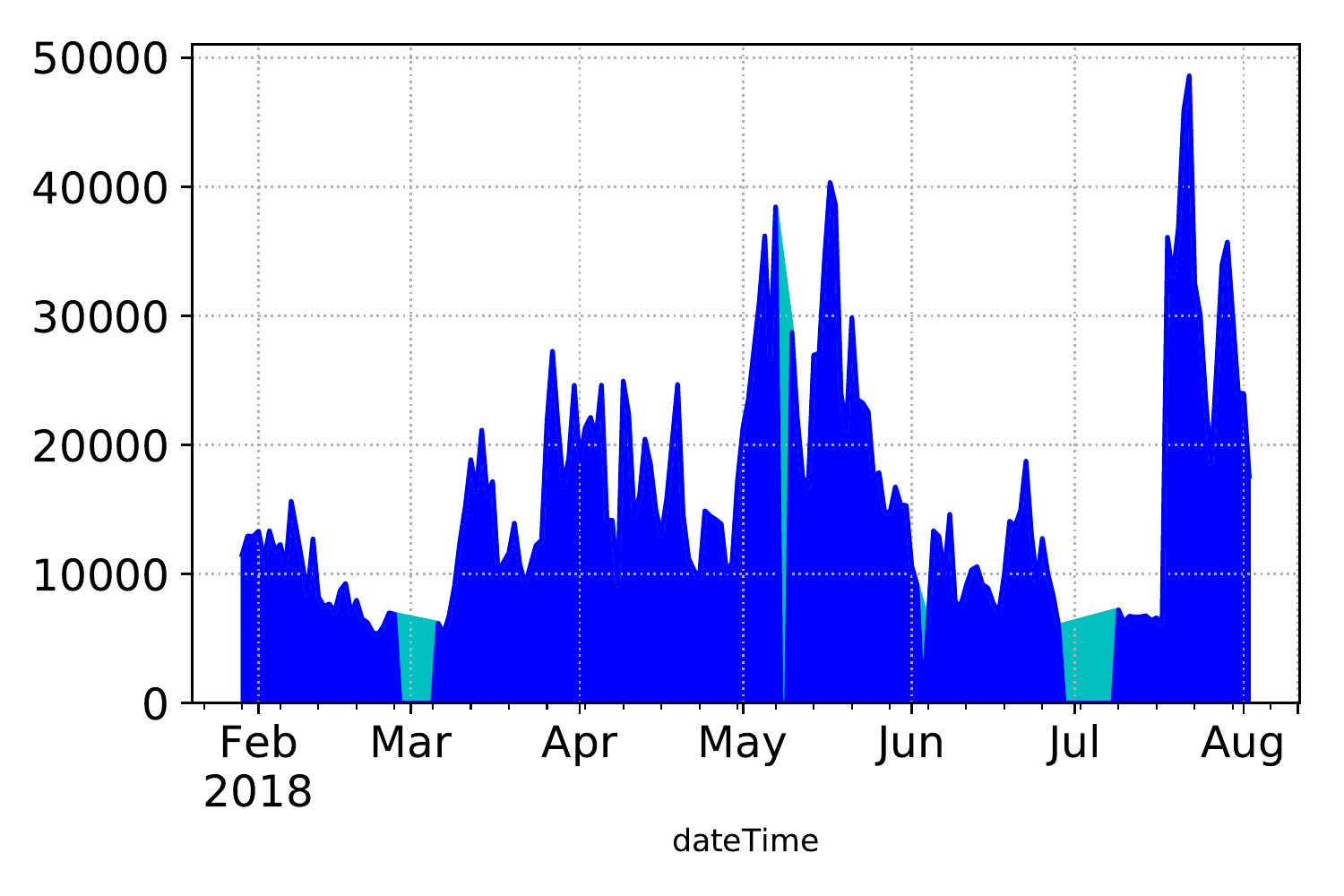}\label{fig:daily_users}
    }
        \subfigure[Distribution of i) number of days each user was observed accessing mining domains and ii) the total number of transactions per user.]{
        \includegraphics[width=0.31\textwidth]{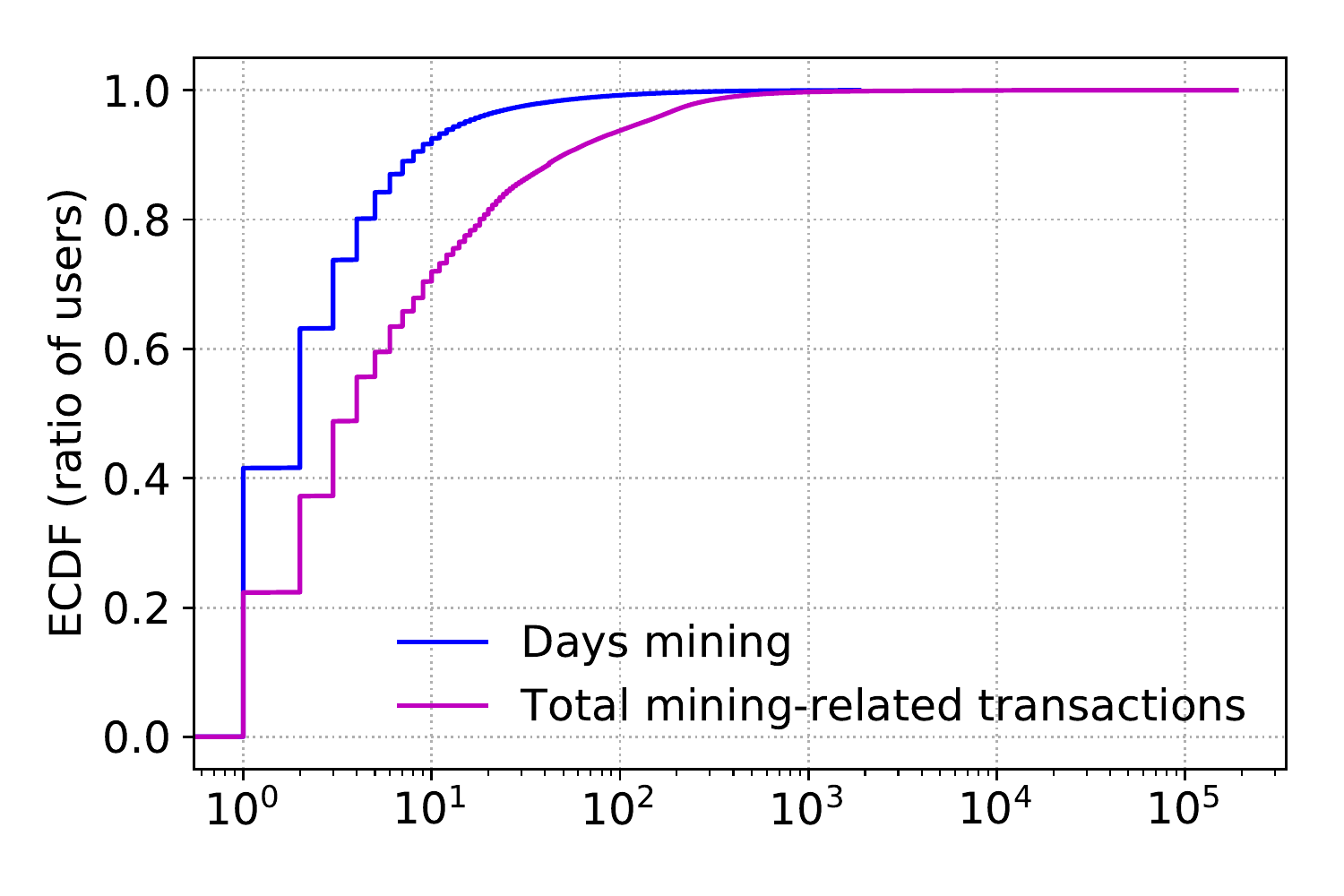}\label{fig:days_active}
    }
         \subfigure[Per-model percentage of active devices that accessed mining domains during 6 months. For comparison we also include the percentage for all Apple and Android devices.]{
        \includegraphics[width=0.31\textwidth]{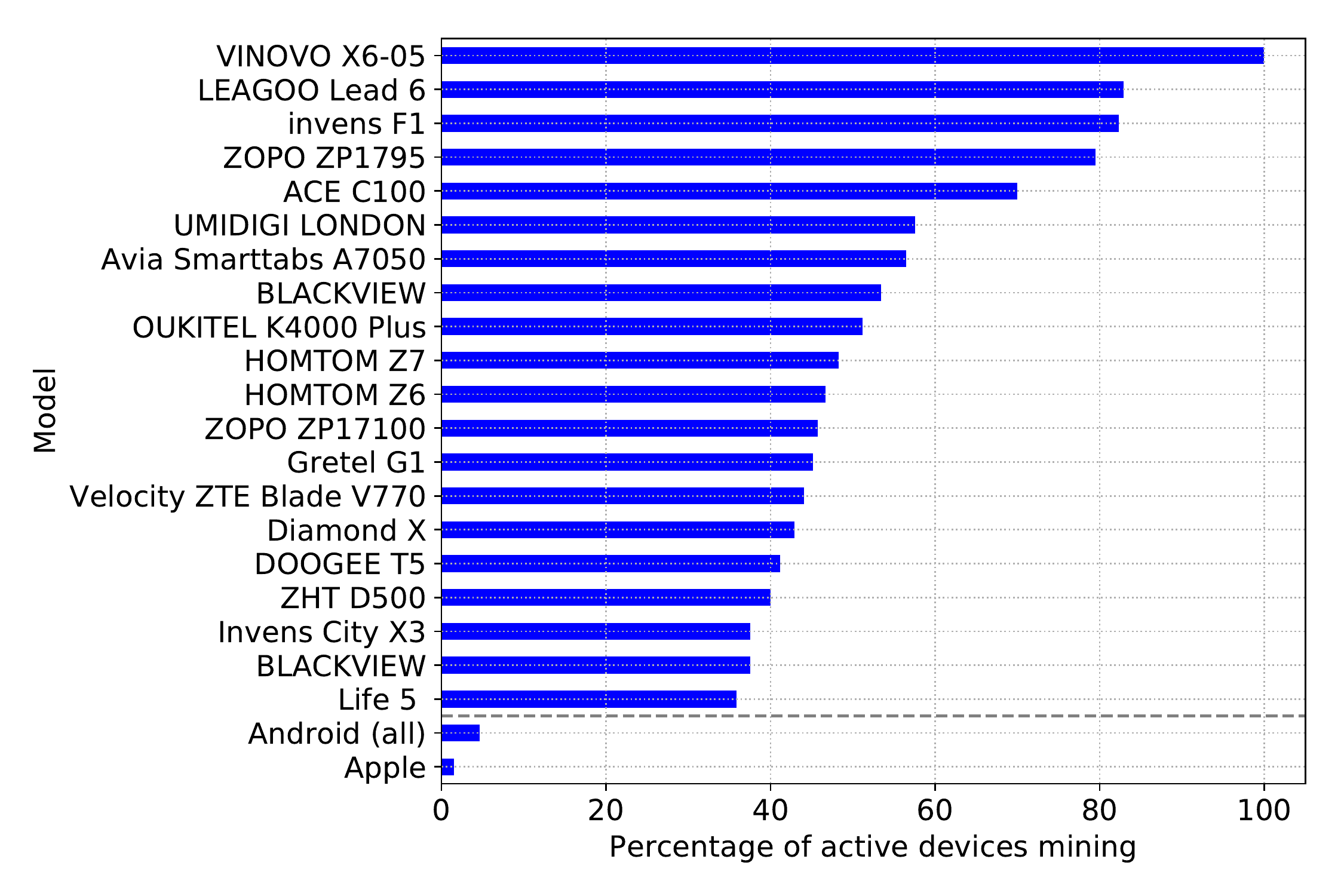}\label{fig:per_mining}
    }
    \vspace{-1.5em}
    \caption{Mobile ISP analysis.}
    \label{fig:mobile-isp-analysis}
    \vspace{-1em}
    \end{center}
\end{figure*}

The data from our mobile ISP covers a much larger
user base and a long observation window (half a year). The picture that our
data establishes is entirely consistent with our network monitoring of research
networks. We identify 91m transactions over six months as mining-related. Of
tens of millions of subscribers, only 1.3m unique users are ever affected.
Coinhive dominates: we identify 30m transactions as access to \coinhive (860k
unique users).  \Cref{fig:daily_users} shows the daily number of unique users
who access domains that we have reason to suspect of hosting mining code.  On
average, 
15k devices access these per day (out of tens of millions of active devices we
monitor). The vast majority (97\%) of transactions relate to \coinhive.

We identify the same peaks in 2018-04 and 2018-05 in this data source,
coinciding with the Drupal vulnerability.  A sudden peak at the end of 2018-07
occurs only in this data set, however. We have no compelling theory why this
peak appears only in this data source.  Inspecting the transactions manually,
we confirm that they constitute increased download activity of the \coinhive
script.

\mysubpara{Frequency of user exposure} Mining on resource-constrained mobile
devices is more detrimental to the browsing experience than mining on desktop
computers. We are interested how long users are exposed, given that the authors
of~\cite{Hong2018} claim that deployed mining code remains on a site for
several weeks. \Cref{fig:days_active} shows the number of transactions and days
that users access mining-related domains. Most users access such domains fewer
than four times over six months.  There is a long tail (0.5\% of users) who
have more than 100 mining-related transactions. We also observe that most users
are only affected for a couple of days over the entire period of six months.
Surprisingly, however, some users seem to access mining domains very regularly: 
90k users do so for seven or more days and 
14k access the domains for 90 days or more. As we show now, this is because
of a previously unknown attack vector.

\mysubpara{Unknown injection vector} \Cref{fig:per_mining} plots the percentage
of actively mining devices per model. We compute this by normalizing the number
of unique devices accessing mining domains with the total number of such
devices in the network over the same period. We only consider models with at
least 50 unique active devices. We identify certain Android devices that are
much more prominent than others. \textit{All} 56 Vinovo X6-05 devices active
during this period access mining websites. 136 out of 164 Leagoo Lead 6 devices
and 353 out of 613 Umidigi London devices do so as well. Statistically, this is
quite unlikely, and hence we hypothesize that some process other than normal
user behavior causes the access. We survey the Amazon product pages for the
devices and find that although they are very affordable, they are also equipped
with relatively powerful hardware. For example, the Umidigi London is a
quad-core, 5-inch smartphone with 1GB of RAM that cost only \$US\,75 in 2018.
Interestingly, many buyers complain in reviews about advertisements appearing on
the home screen. Others notice background activity, especially when charging,
or applications that are installed without user interaction.

We buy one such device for testing purposes. A few weeks after activation, the
phone starts displaying pop-ups, advertisements, and various apps are installed
without user intervention. Google recently reported the Triada malware to have
been injected into the supply chain for some of our affected
devices~\cite{MalwarePhones2019}. Triada is known to cause similar behaviors.
We test for the typical signatures identified by Google; however, none of them
are present in our firmware. This suggests that our device is controlled in a
different but possibly related way. It may well be that the apps installed
without user permission load web sites in the background that contain mining
code.  It is unclear if this is the intention behind the installed apps or only
a side-effect.\footnote{We make the firmware of our device available to
interested researchers.}

\subsection{Alternative Attack Vectors: Tampering with Connections}
\label{sec:sub:results-open-proxies-tor}

In line with our previous findings, we find few incidences of open proxies or
Tor exit nodes tampering with traffic for the purpose of mining cryptocurrency.

\begin{figure}[tb]
    \centering
    \includegraphics[width=2.3in]{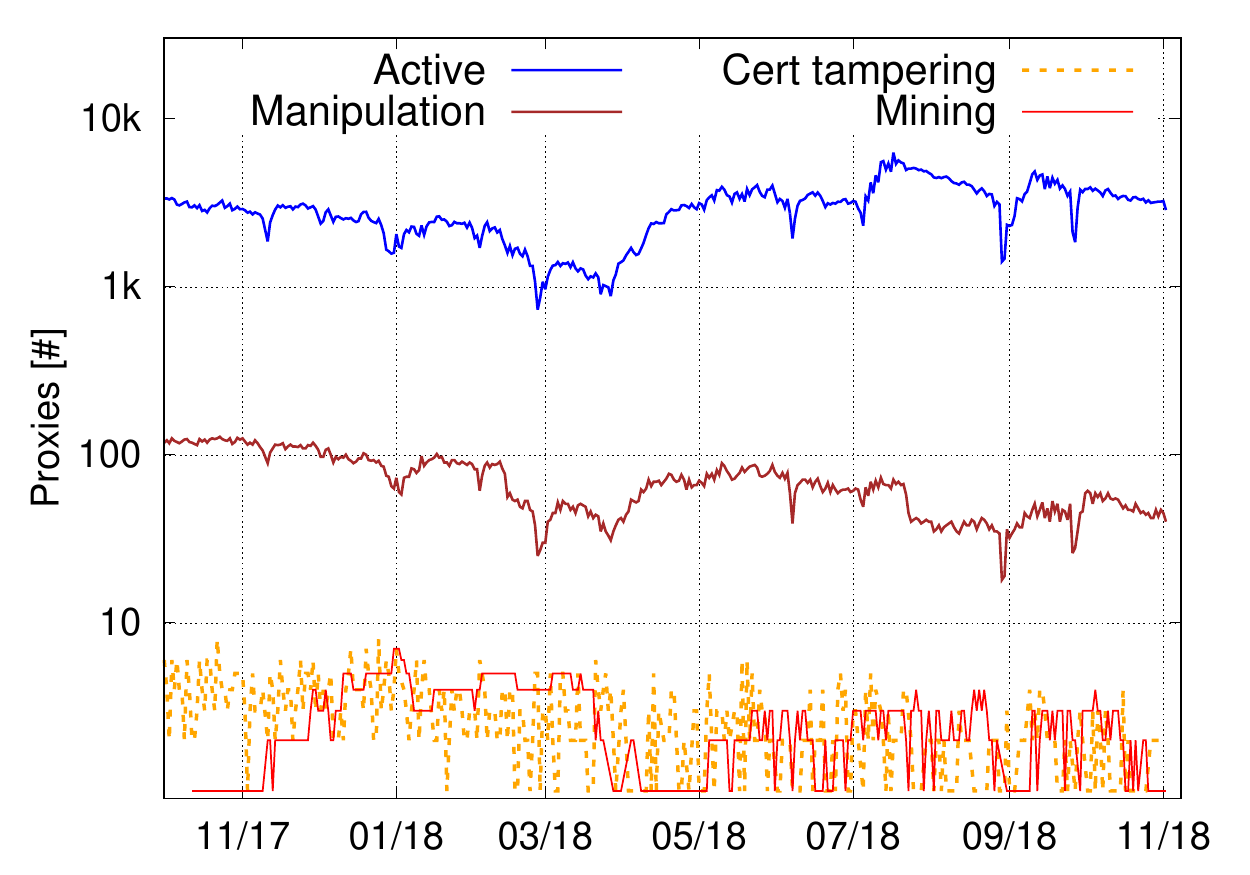}
    \caption{Open proxy ecosystem and miners evolution.}
    \vspace{-1.3em}
    \label{fig:proxies_time}
\end{figure}

\mysubpara{Open proxies} Out of \num{15892} proxies active and fast enough to
deliver a site within our 60 seconds timeout, only 282 (1.7\%) modify content.
6\% of these proxies inject mining code, \ie a total of 0.11\% of the tested
proxies. We verify manually that each case is a true positive. All injections
are related to \coinhive. Figure~\ref{fig:proxies_time} shows our results over
time.  We see most mining activity from 2017-12 to 2018-03, which coincides
with the ramp-up phase of in-browser mining and higher Monero prices. We never
find more than ten proxies injecting mining code at any time. As in our other
data sources, we see a decline beginning in mid-2018. We track whether proxies
also manipulate TLS certificates, \ie stage a Person-in-the-middle attack on
HTTPS connections.  We find a similar fraction of proxies doing this---but when
we inspect these cases, they are never the same proxies that inject mining code
into HTTP.

We measure how long proxies attempt code injection and establish a median of 22
days. A similar number has been reported for other malicious tampering by open
proxies in~\cite{proxytorrent}. It is also close to the number 
reported in~\cite{Hong2018} for websites hosting mining code: a third stops the
mining within two weeks.  The short duration during which injection of code by
proxies happens contrasts with the overall lifetime of proxies, which spans
months. One proxy, for example, is active for about one year but only injects
mining code for one month.

\mysubpara{Injection in Tor} Our analysis of injections at Tor exit nodes spans
more than 200 days, with an average of 680 successful downloads of the bait
page per day. The number of exit nodes in our measurement period is between 900
and 1100, although not every exit node provides us with a result. This is due
to the way how onion circuits are created and the fact that exit nodes set
policies for exit traffic. We exclude exit nodes that do not forward our HTTP
requests. Although we use a very broad range of keywords, we find only one Tor
exit node injecting mining code. This specific exit node injects an
unobfuscated \coinhive script.  It is only active for 13 days, during which we
see the injection on four days. It is instructive to compare this with previous
results on content modification in exit nodes. Winter \etal \cite{Winter2014}
found significant malicious behavior among Tor exit nodes in 2014. The number
of exit nodes has not changed significantly since then, but Winter \etal found
40 exit nodes tampering with HTTPS connections and 27 exit nodes stealing
credentials from HTTP connections. However, Tor has a tech-savvy user base
likely to notice tampering. The network now encourages users to report
malicious exit nodes, which can be flagged in the directory service. We believe
that Tor was hence never particularly appealing to attackers.

\subsection{Understanding User Visits to Mining Sites}
\label{sec:sub:results-tracking} We investigate the extent of mining
experienced by users of our browser extension. Note that it is configured in
such a way that any mining code we find must come from the visited site as
users cannot access malicious proxies.

We updated and extended the classifiers over time to ensure coverage and
accuracy. Table~\ref{tab:ciao-over-time} summarizes our findings. Once again,
we find that users are very occasionally affected by mining. Mining sites are
rarely accessed by users: they account for just 0.03\% of the total number of
domains our users visited. We find the characteristic bumps in \mbox{2018-04} and
\mbox{2018-05}, which coincide with the attacks on the Drupal content management
system. This corroborates our previous results from passive measurement.  We
notice a peak in \mbox{2018-07}. However, closer inspection reveals that this is an
automated tool using our plugin repeatedly to visit a miner domain. Such tools
have also been reported in~\cite{proxytorrent}. The peak is hence not related
to the one we find in our ISP data. \coinhive is the most common mining pool we
encounter---we find it in 93\% of sessions. This is followed by \authedmine,
\jsecoin, Cryptoloot, and Coinimp, attesting to the growing significance of
alternative providers.

We investigate the duration of HTTP sessions with and without mining taking
place.  The median grows from six seconds for regular sessions to 30 seconds
when mining occurs. This seems counter-intuitive as one would expect users to
be annoyed by high CPU usage. However, it is likely that some mining sites
offer attractive content, as also suggested in~\cite{Rueth2018,Bijmans2019}
(entertainment and adult content). However, most sessions are too short to
contribute significant mining results to a mining pool: just 10\% of mining
sessions last more than 16 minutes.

\footnotesize{
\begin{table}[tb]
	    \centering
        \caption{Visits to mining sites identified by our browser extension.}
    \label{tab:ciao-over-time}
    \vspace{-0.5em}
    \scalebox{0.8}{
    \begin{tabularx}{0.95\columnwidth}{X r r r}
        \toprule
        Month (2018) & Total visits & Visits to mining sites & \% \\
        \midrule
        Feb & \num{113257} & 18 &  0.016\%\\ 
                Mar  & \num{206518} & 33 &   0.016\%\\
                Apr & \num{171555} & 89 & 0.052\% \\
                May & \num{153256} & 53 & 0.034\% \\
                Jun & \num{195092} & 40 & 0.021\% \\
                (Jul) & (\num{160373}) & (461) & (0.287)\% \\
        \bottomrule
    \end{tabularx}
}
    \end{table}
}\normalsize

\section{Discussion and Conclusion}
\label{sec:discussion}

In this paper, we set out to analyze the exposure of Web users to
cryptocurrency mining.  In a nutshell, do reported deployment rates between
0.01--0.1\% imply a great risk for users or not? The conclusion that we offer
is that the risk to users was greatly overestimated. This is due to the nearly
exclusive use of active scans in previous work. In contrast to previous work,
our study is \emph{retrospective} and \emph{longitudinal} in nature: we pool
datasets raised by several groups around the planet to paint a more
comprehensive picture of user exposure. In the following, we combine and
discuss the results from our various data sources in context.

\mysubpara{Exposure was very low} All our measurements show users were very
rarely exposed to mining. We note that we verified high coverage and accuracy
for our classifiers. Our own active scans, which provide some input to our
passive observations, are consistent with related work. An important take-away
to consider in future studies of Web attacks is the degree to which the use of
deployment numbers alone is a very insufficient metric to express risk to
users.  The long-tail distribution of browsing preferences must be accounted
for to quantify risks meaningfully. In our data, we were able to establish the
long-tail distribution for mining sites that users actually connect to. Taking
this into account, we believe that blocklists in browsers or on network
gateways were actually also more effective than previous work gave them credit
for; they should not be dismissed.

\mysubpara{Exposure was very short} The number of connections to mining proxies
alone do not reveal how users react to mining sites, and whether there is any
measurable contribution to a pool's overall mining. The data we obtain from our
browser extension gives us a unique perspective here. It corroborates that
users encounter mining very rarely but also shows that most users stay briefly
if they come across mining sites: usually around 30 seconds. As revenue is a
driving force in the operations of cybercriminals, it is important to take this
into account when new, different attacks exploit users' CPU time.

\mysubpara{Risk from software monoculture} The peaks in our passive
measurements correspond to compromises of a widely used content management
system. This should be taken as yet another warning that software monocultures,
combined with the known slow update rates in the web ecosystem, put users at
risk too easily.

\mysubpara{Uncommon attack vectors were tried} We expose one attack vector that
was hitherto unknown: mobile devices loading apps that lead the devices to
visit mining sites, possibly in the background. While a relatively infrequent
occurence, our findings support the call for more stringent checks of operating
systems, including in supply chains.

\section*{Acknowledgements}

This work was supported partially by the Australian Government through the
Australian Research Councils' Discovery Projects Scheme (project DP180104030).
The views expressed herein are those of the authors and are not necessarily
those of the Australian Government or Australian Research Council. Ralph Holz
received partial funding from an Early Career Researcher Award by the Faculty
of Engineering, University of Sydney. This work was supported by the US
National Science Foundation under award CNS-1528156. Any opinions, findings,
and conclusions or recommendations expressed in this material are those of the
authors or originators, and do not necessarily reflect the views of the NSF.

\bibliographystyle{IEEEtran}
\bibliography{biblio-new}

\end{document}